\documentclass[pre,preprint]{revtex4}

\usepackage{graphicx}
\usepackage{amsmath,amssymb}
\begin{document}

\title{Random walks, diffusion limited aggregation in a wedge, and average conformal maps}
\author{Leonard M. Sander}
\email{lsander@umich.edu}
\affiliation{Michigan Center for Theoretical Physics, Department of Physics,
	University of Michigan, Ann Arbor, Michigan, 48109-1120}

\author{Ell\'ak Somfai}
\email{ellak@lorentz.leidenuniv.nl}
\affiliation{Instituut-Lorentz, University of Leiden, P. O. Box 9506, 2300 RA Leiden, Netherlands}

\date{\today}
\begin{abstract}
We investigate diffusion-limited aggregation (DLA) in a wedge geometry. Arneodo and collaborators have suggested that the ensemble average of DLA cluster density should be close to the noise-free selected Saffman-Taylor finger. We show that a different, but related,  ensemble average, that of the conformal maps associated with  random clusters, yields a non-trivial shape which is also not far from the  Saffman-Taylor finger.  However, we have previously demonstrated that the same average of DLA in a channel geometry is \emph{not} the Saffman-Taylor finger.  This casts doubt on the idea that the average of noisy diffusion-limited growth is governed by a simple transcription of noise-free results.  
 \end{abstract}
\maketitle 
 \section*{}{\bf
\noindent The diffusion-limited aggregation model uses aggregating random walkers to form a random fractal pattern. We can also use random walkers to generate a conformal map from the exterior of the cluster  to the exterior of the unit circle. The map contains all the information about the growth probabilities for various points on the cluster. We define the pattern generated by the ensemble average of the conformal maps as a kind of mean-field pattern for DLA. In previous work we showed that there seemed to be an intriguing relationship between DLA in a channel and the noise-free Saffman-Taylor finger. We show here that this relationship is much more ambiguous when we apply it  to DLA in a wedge geometry. This result casts doubt on the `averaging conjecture' which holds that the average of noisy growth `remembers' noise-free results.
 }
 
 \section{Introduction}
The last century has abounded with examples of unexpected richness in the problem of brownian motion as formulated by Einstein in his seminal paper in 1905 \cite{einstein05}. In the same year,  Pearson \cite{pearson05} pictured the process as a \emph{random walk}. In the past century random walks and brownian motion have become  central themes of statistical physics. The problem is particularly astonishing in that it constantly generates new ways to think about nature, and new descriptions of physical processes. One example was the discovery by Witten and Sander \cite{witten81,witten83} that aggregating random walkers give rise to random fractal patterns. This process, diffusion-limited aggregation (DLA) is simple to formulate: a seed particle is put at a point, and then a random walking particle is launched and allowed to proceed until it touches the seed; then it stops. Then another walker diffuses until it touches one of the first two, and so on. We study the cluster generated. This is the simplest paradigm of noise-dominated growth limited by diffusion, a common natural process.
 
This process has been studied intensively over the last twenty-three years (for a review, see \cite{sander00}), but there is still not complete theoretical understanding, though some recent progress has been made \cite{ball02prl} in describing local correlations. In this paper we investigate a property of the \emph{ensemble} of DLA clusters namely the generation of \emph{average shapes}. Our motivation comes from the remarkable suggestion of Arneodo and collaborators \cite{arneodo89,arneodo96} that the (somewhat arbitrarily defined) ensemble average shape of noisy DLA clusters, would be the pattern generated by {noise-free} diffusion-limited growth. We will refer to this  as the \emph{averaging conjecture}. The relevant case of noise-free growth is the Saffman-Taylor  viscous finger \cite{saffman58,pelce04}, i.e, the shape of the surface of an inviscid fluid invading a viscous one, as water into oil. The work of Arneodo, et al. showed that Saffman-Taylor fingers in a channel and a wedge were close, but not exactly the same, as  DLA averages.  However the work had a number of arbitrary parameters, and, in the wedge, there were serious ambiguities.

We investigated the averaging conjecture in our work \cite{somfai03} on DLA in a channel with reflecting boundaries at the walls. There, we formulated a new definition of the average shape by averaging the \emph{conformal map} \cite{hastings98,davidovitch99,somfai99}  that generates the cluster -- this amounts to weighting points on the surface according to their growth probability. In that work we found that the average shape of DLA (using our definition) was \emph{not} a Saffman-Taylor viscous finger.

However, a number of authors \cite{ball02prl,barra01} had already questioned whether DLA and viscous fingers are actually closely related growth processes;  DLA does not have surface tension like viscous fingering, but rather a fixed particle size that defines the tip radius. We followed up the suggestion \cite{ball02prl,ball03pre} that fluid flow with surface tension \emph{is} closely related to a variant of DLA called the dielectric breakdown model (DBM) \cite{niemeyer84} with parameter $\eta \approx 1.2$ (to be defined below). In fact, the suitably averaged DBM clusters then turned out to fit the Saffman-Taylor shape quite closely. 

Here we look at the related problem of DLA in a wedge. Previously \cite{kessler98} we have shown that tip-splitting in this geometry gives us access to local correlations. In the present context of looking at average shapes, the wedge is interesting in several respects. This is a richer problem than that of the channel in that the wedge angle is a free parameter, and the shape of the fluid invasion is more complex \cite{thome89,tu91,benamar91,benamar91b}. We pose the question of whether the DBM shape in a wedge is also a good approximation to the Saffman-Taylor finger in a wedge. 

This study has a special significance in this focus issue on brownian motion. Our work here is primarily numerical. The simulations are \emph{all} done by random walker sampling: we generate DLA clusters, DBM clusters, and conformal maps by this single method which, as it happens, is by far the most efficient method available. This century-old technique has not lost its freshness and power.

\section{DLA and DBM in a wedge}
There are now available very sophisticated schemes for generating DLA clusters. The one we use is based on the method of hierarchical maps \cite{ball85}. In this method space is divided into regions of various sizes which help keep track of the nearest points on the cluster. Then the random walker can make large jumps in empty regions, vastly speeding up the computation. With this method the time to create an $N$ particle cluster is proportional to $N^p$ with $p\approx 1.1$. 

\begin{figure}
\begin{center}
\includegraphics[clip,height=80mm]{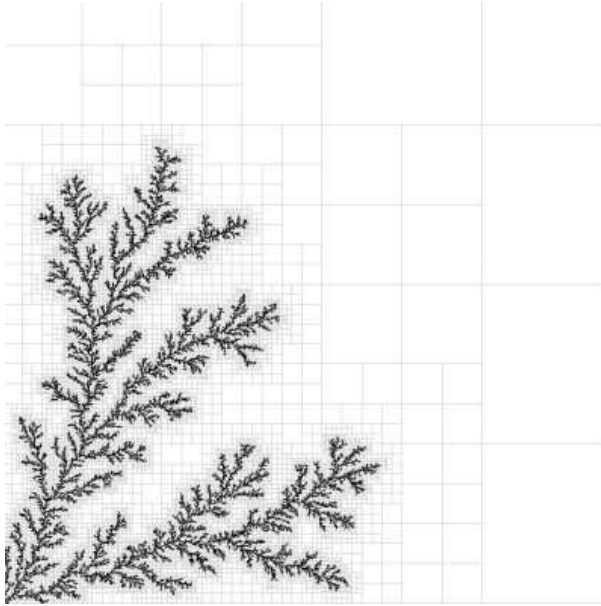}
\caption{A DLA cluster in a 90$^\circ$ wedge with reflecting boundaries. The structure of the hierarchical maps is also shown.}
\label{dla90}
\end{center}
\end{figure}

We need to make DLA clusters in a wedge with reflecting boundary conditions. We do this by means of a trick. It there is a wall at some position, every time we deposit a walker, we deposit an image walker reflected in the wall. Suppose we are interested in 90$^\circ$ wedges. Then we have two perpendicular walls, and four walkers are deposited at once. Using this method we can use a radial DLA code to produce wedges of opening angles $180^\circ/n, n=1,2,...$. In this paper we will concentrate on 90$^\circ$ and 60$^\circ$ wedges. An example of a 90$^\circ$ wedge is shown in Figure~\ref{dla90}.

In the following we will see that we need to consider DBM clusters. These are defined as follows: we imagine that the cluster is a grounded conductor with unit charge. Outside the cluster define a potential, $\phi$ so that:
\begin{equation}
\label{dbm1} 
\nabla^2 \phi =0, \quad \phi_s=0.
\end{equation} 
This defines an `electric field' on the surface, $\partial \phi/\partial n_s$. Then the dielectric breakdown model takes the growth velocity on the surface of the cluster to be:
\begin{equation}
\label{dbm2}
v_n \propto \left[\frac{\partial \phi}{\partial n} \right]^{\eta}_s.
\end{equation} 
We interpret this equation probabilistically: $v_n$ is taken as the density of the growth probability  on the surface, $\mu$. In practice, add a particle at a point on the cluster with probability proportional to $v_n$.
 It is known that at large scales DBM clusters with $\eta=1$ have the same scaling as DLA. 
 
The original method  \cite{niemeyer84} to grow DBM clusters was to solve Eq.~(\ref{dbm1}) by relaxation. This algorithm is very slow, and is not practical for generating large clusters. Recently \cite{somfai04} we have introduced a method of growing DBM clusters by random walker sampling. The key to this method is to define the age, $a_1$ of a growth site. This is the number of random walkers that have landed anywhere on the cluster since the last particle grew at the site. We can also define $a_k$, the number of walker that have landed since the $k^{th}$ most recent particle grew there.

Since the frequency of landing of random walkers is proportional to $\mu$, it is clear that $1/a_1, 2/a_2, ..$ at a site are estimates of $\mu$ at that site. We have shown \cite{somfai04} that this estimate is adequate to allow us to grow DBM clusters. In our work here we use $a_3$ to estimate the probability. 

\begin{figure}
\begin{center}
\includegraphics[clip,height=60mm]{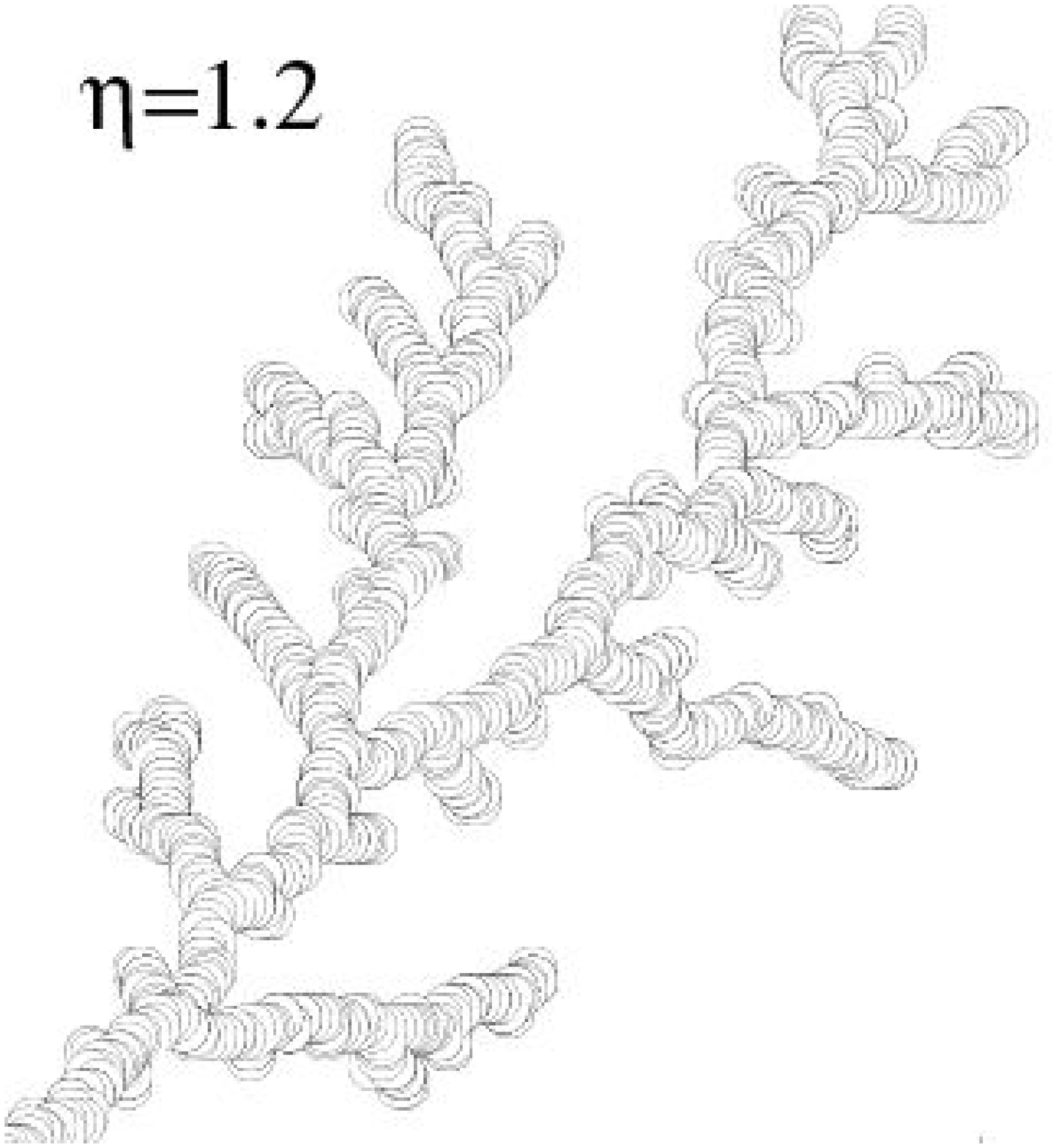}\hspace{10mm}
\includegraphics[clip,height=60mm]{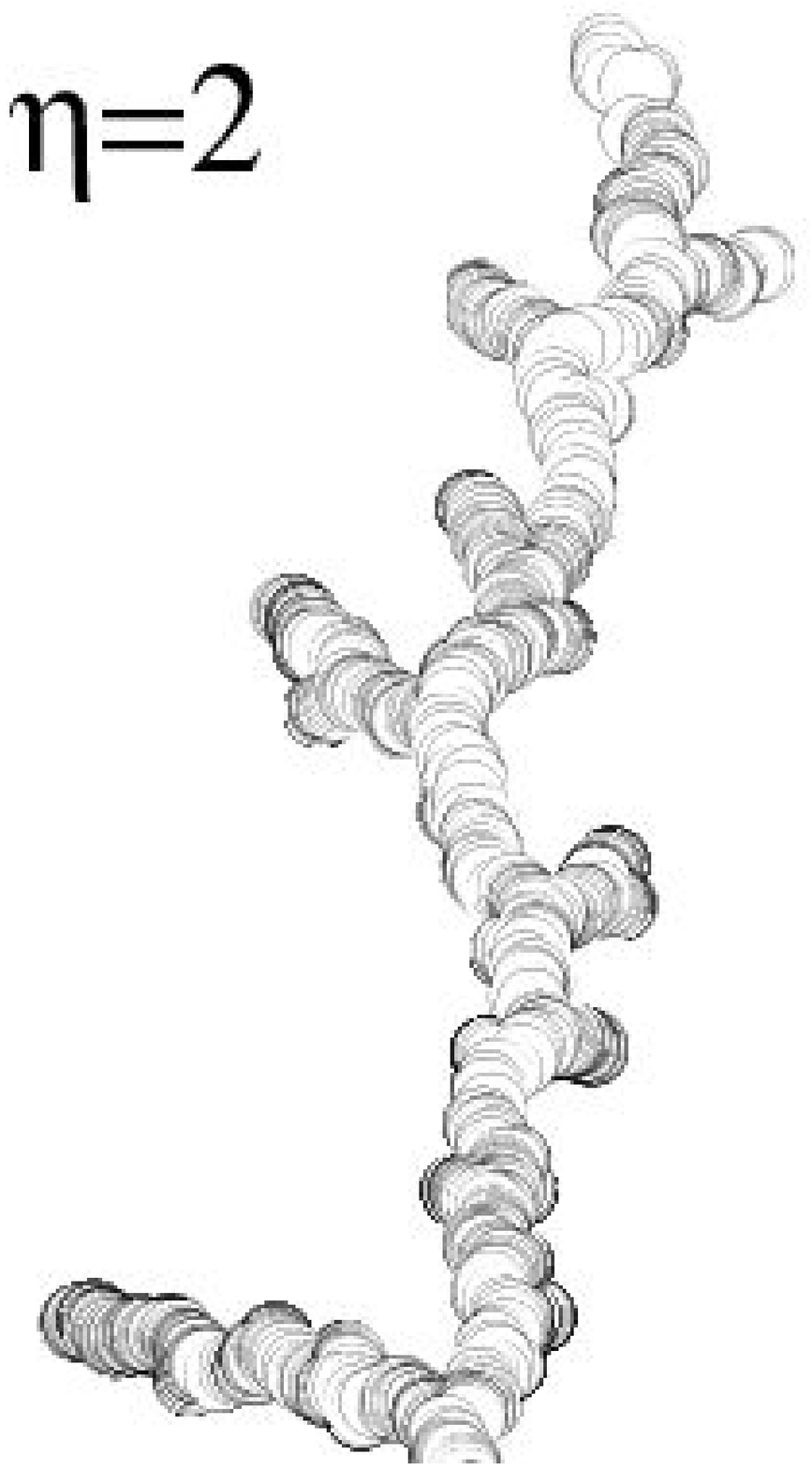}
\caption{A portion of DBM clusters for two different values of $\eta$. For $\eta=2$ the overlapping of particles and the enhanced growth at the tips  is easy to see. }
\label{eta}
\end{center}
\end{figure}

The method of growth is as follows: if a particle lands at a site with low probability, we arrange to have it add little to the cluster, and at high probability sites, add a good deal.  This is accomplished by adding a mass, $\delta m =A \mu^{\eta-1} \propto a_n^{1-\eta}$. 
(The power is $\eta-1$ because we already have a probability $\mu$ for the walker to land at the site.)
In practice, when a particle is added at a site, it is moved onto the existing particle so that a portion proportional to $a_k^{1-\eta}$ contributes to new growth. We also change the prefactor, $A$, as we go along to make an efficient code. For details, see \cite{somfai04}.  Examples of this is shown in Fig.~(\ref{eta}).

 The computations we discuss below are averages over an ensemble of off-lattice DLA and DBM clusters grown in this way. Our DLA clusters had $1,000,000$ particles in the wedge and we averaged over an ensemble of   400 realizations. For the DBM clusters we had $500,000$ particles in the 60$^\circ$ wedge, $1,000,000$ particles in the 90$^\circ$ wedge and also 400 realizations. Our motivation for going to these large sizes is the fact that DLA suffers slow crossovers \cite{somfai99,ball02pre}. If we use small clusters we are not seeing the asymptotic behavior.

\section{Conformal maps}
In recent years a number of groups  \cite{hastings98,davidovitch99} have looked at DLA in an entirely new way. The cluster shape is considered to be a grounded conductor, as above, and the complex potential, $\phi$,  is sought, as in Eq.~(\ref{dbm1}). The technique introduced was to define a conformal map from the exterior of the cluster to the exterior of the unit circle. The  Laplace equation, Eq.~(\ref{dbm1}) can be solved easily  outside the circle, and the solution mapped back to the cluster. Of course, the solution for  $\partial \phi/\partial n$ is uniform on the circle. Thus the image of two parts of the cluster perimeter with the same growth probability will map to parts of the unit circle with the same length. Put another way, the inverse images of uniformly spaced points on the unit circle are distributed on the cluster with density $\mu$. 

In order to construct the map Hastings and Levitov \cite{hastings98} invented an iterative technique which grows a cluster and calculates the map at the same time. This is a practical method, but slow. In Ref.~\cite{somfai99} we constructed an alternative method which is much faster. We grow a cluster by the  conventional fast scheme using random walkers. Then we freeze the cluster at the desired size, launch $n$ random walkers as probes, and record where they land. Then the values of the map on the unit circle are found as follows. We choose one of landing positions as a starting point, say along the $x$-axis, and number the walkers around the perimeter of the cluster. By the observation above, if we are at walker $m$, we know that we have turned an angle $\theta = 2\pi m/n + \mathcal{O}(n^{-1/2})$  from the image of the $x$-axis on the unit circle. In Ref. \cite{somfai99} we used this information, which samples the boundary values of the conformal map, to construct the map itself by analytic continuation.

\begin{figure}
\begin{center}
\includegraphics[clip,height=80mm]{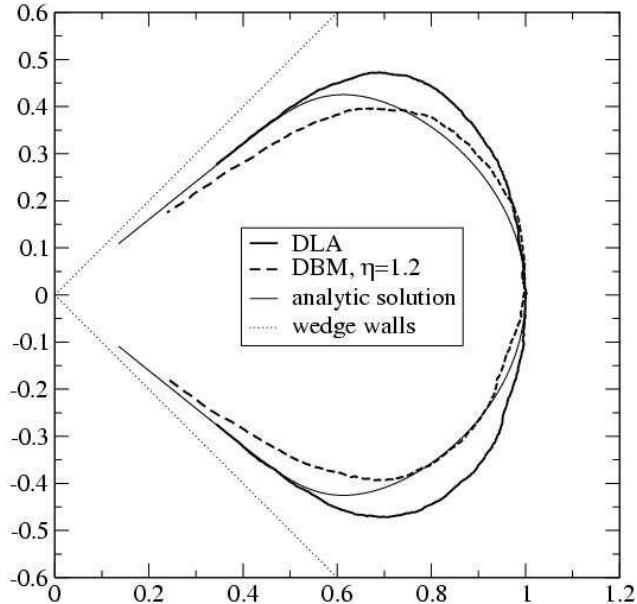}
\caption{The averaged profile of DLA and DBM clusters with $\eta=1.2$ in a 90$^\circ$ wedge. Also shown is the  the analytic solution for a Saffman-Taylor finger with the selected $\lambda$ from \cite{thome89,tu91}.}
\label{90}
\end{center}
\end{figure}

\begin{figure}
\begin{center}
\includegraphics[clip,height=80mm]{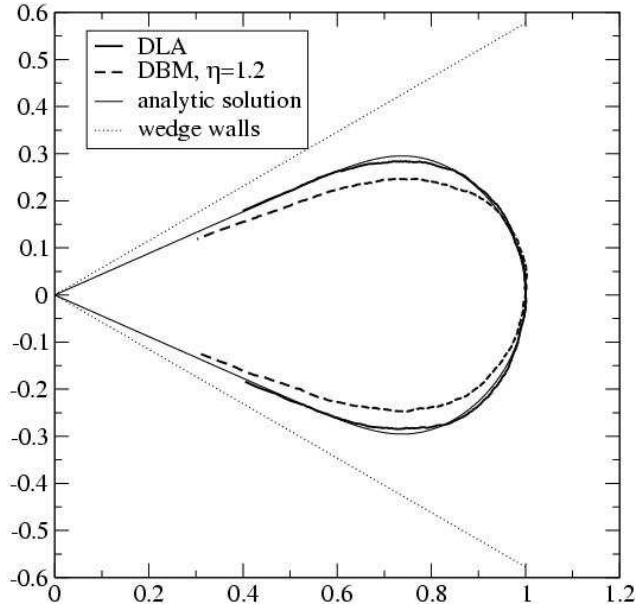}
\caption{The averaged profile of DLA and DBM clusters with $\eta=1.2$ in a 60$^\circ$ wedge. Also shown is the  the analytic solution for a Saffman-Taylor finger with the selected $\lambda$ from \cite{thome89,tu91}.}
\label{60}
\end{center}
\end{figure}

For our purposes here we use the information in a simpler way. We constructed the map to the unit circle using $n=100,000$ for each cluster that we grew. For each member of the ensemble there is a point, $\mathbf{r}(\theta)$ whose image is a point on the unit circle at $e^{i\theta}$. Our definition of the ensemble average shape generated by the DLA or DBM process \cite{somfai04} is the ensemble average of $\mathbf{r}(\theta)$, i.e., the centroid of those points. In  Figures~(\ref{90}) and (\ref{60}),  below we show the average shapes in  90$^\circ$ and 60$^circ$ wedges.

This definition has a number of advantages: it is unambiguous, in contrast to definitions based on density profiles (see below). It is an average which is weighted by probability. That is, we are sampling where the growing tips of the cluster are located. It might be accessible to theoretical investigation since the growth process is the defining property of DLA.

\section{The averaging conjecture and mean-field theory}
\subsection{Saffman-Taylor fingers}
The investigation of Hele-Shaw flow in a channel is an old subject and a good deal is known about it. In particular, the invasion of a viscous fluid by an inviscid one is one of the most famous of all pattern formation problems \cite{saffman58,pelce04}. The general result is that the inviscid fluid forms a finger, the Saffman-Taylor finger, which fills a fraction, $\lambda$, of the channel. 
For each $\lambda$ in $[0,1]$ there is a solution to the fluid-flow equations without surface tension.  For small surface tension, the pattern is very close to that with zero surface tension and  $\lambda=1/2$. This is called the selected value. The finger elongates in time, and is of constant shape in a moving reference frame.

The theory of this sort of flow is remarkably well developed. A few basic notions are necessary here:  the fluid velocity, $\mathbf{v}$,  is derivable from a potential in this case, and is governed by D'Arcy's law. The upshot of these two facts is that:
\begin{equation}
\label{heleshaw}
\nabla \cdot \mathbf{v} = 0 = \nabla^2 \phi; \quad  v_n \propto \left[\frac{\partial \phi}{\partial n} \right]_s.
\end{equation}   
 That is, fluid flow obeys the same equations as DBM (cf., Eq.~(\ref{dbm2})) with $\eta=1$ \cite{paterson84}, so that the Saffman-Taylor finger has a good deal in common with DLA. The bubble of inviscid fluid plays the role of the aggregate. However, there are two salient differences: DLA does not have surface tension but, rather, a finite size cutoff; and DLA is dominated by noise, whereas the Saffman-Taylor finger is a stable, noise-free pattern that is observed for slow flows in a channel.
 
 There is another problem related to the channel problem which also admits of an exact solution, that of viscous fingering in a wedge. In this case, for short times there is also a selected shape in experiments, at least for a finite time \cite{thome89}.  Tu and Ben Amar \cite{tu91, benamar91, benamar91b} worked out the theory in this case and showed that there is a self-similar shape that is selected. It is a non-trivial pattern whose form is given by a differential equation that needs to be solved numerically. Associated with the problem is a selected angle, defined as the opening angle at the base of the wedge; see Figures~(\ref{90}) and (\ref{60}). The ratio of the opening angle of the finger to that of the wedge is also called $\lambda$. With no surface tension there is a solution for all $\lambda$, but, once more, there is a selected value which depends on the wedge angle. There is a complication in the fluid-flow problem. If the inviscid fluid is pumped in a constant pressure, as time goes on there will always come a point when the pattern is unstable against tip-splitting. 
 
\subsection{Averaging}
Arneodo and collaborators \cite{arneodo89,arneodo96}  exploited the resemblance between DLA growth and viscous fingering in the following way: they speculated that the average of many DLA clusters would, in some sense, remove the noise, and recover the noise-free pattern. Since DLA has no surface tension, but rather a fixed particle size (playing, roughly, the role of the capillary length) they assumed that the limit of small surface tension was the appropriate one. This is what we refer to as the averaging conjecture.

They tested the conjecture by generating on-lattice DLA clusters in a channel and averaging the density, point by point \cite{arneodo89}. The density average, 
$ \rho(\mathbf{r})$ is a  function that goes to zero at the edge of the channel, and has a maximum, $\rho_{max}$ at the center.  One of the level sets of this function, $\mathbf{r_s}: \{\rho(\mathbf{r_s})= \frac{1}{2}\rho_{max}\}$ traced out a Saffman-Taylor finger with $\lambda=1/2$. 

However, closer scrutiny made the picture more complex. Lattice effects are known to distort DLA clusters, and are irrelevant to the kind of physics being considered. Therefore, they returned to the problem \cite{arneodo96} and generated \emph{off-lattice} DLA clusters. They found that  the  level set at $\frac{1}{2}\rho_{max}$ filled 56\% of the channel rather than 50\%. Or, alternatively, the level set that was needed to make $\lambda=1/2$ was at $0.6 \rho_{max}$. 

For the same problem in wedges of various opening angle, on-lattice DLA density averages once more gave remarkable agreement with selected fingers \cite{arneodo89}. However, off-lattice the situation was different: tip-splitting made the front of the fingers too flat, though there was qualitative agreement with the opening angle of the average density. And, it was necessary to choose a level set somewhat arbitrarily. In fact, since the overall density decreased as the length of the finger increased, it was necessary to define the opening angle by taking a fraction of the maximum density at that distance from the apex of the wedge. 

\subsection{Mean-field theory}
Inspired by the averaging conjecture, Levine, Tu, and collaborators \cite{brener91, levine92} revised the mean-field theory of Witten and Sander \cite{witten83} to attempt to write a proposed set of equations for the mean density of a cluster. They found qualitative agreement with Ref.~\cite{arneodo89}. In Ref.~\cite{arneodo96} the theory was extended, and other work has followed this up more recently \cite{bogoyavlenskiy01}.

The salient features of this theory are that a level set, defined as above, more-or-less fits the channel finger, but that the wedge-based fingers are too flat. 

\subsection{Average conformal maps for DLA in a wedge}
The numerical underpinnings of the averaging conjecture are troublesome in several ways. At the most simple level, the number of particles in the clusters studied were very small, of order $10^3 - 10^4$. We have already remarked that DLA in that regime is far from asymptotic. More significantly, there are far too many fitting parameters in the discussion. The level-set is chosen arbitrarily, and, for many of the discussions, $\lambda$ is chosen to fit the pattern. 

We returned to this problem with our new definition of the ensemble average pattern, described above. In a channel we were able to show \cite{somfai03} that the average shape does not fit any Saffman-Taylor finger. The finger width of the pattern corresponded to $\lambda \approx 0.6$, just as in Ref.~\cite{arneodo96}, despite the different definitions of the average. With the resolution that we had available, we were able to show definitively that the finger we generated  did not fit the Saffman-Taylor pattern for any $\lambda$. 

Here we return to the problem for growth in a wedge. Using the techniques described above, we have grown DLA clusters and averaged for wedge angles of 90$^\circ$ and 60$^\circ$. The results are shown in Figures~(\ref{90}) and (\ref{60}) along with the analytic solutions for the selected finger shape. Now, quite remarkably, the opening angle of the finger does fit rather well to the \emph{selected} $\lambda$. Note that there are no adjustable parameters in this fit whatsoever. The tip of the  finger in the 90$^\circ$ wedge is flatter than it should be. However, for  60$^\circ$ the fit is reasonably good.

\subsection{Average conformal maps for DBM in a wedge}
In our channel work we tried to salvage the averaging conjecture in the following way. There is a theoretical suggestion \cite{ball02prl,ball03pre} that the correct analogy between fluid flow and 
flow with surface tension was not with DLA. Rather, the nature of the short-range cut-off in Hele-Shaw flow is that the tip radius, $R$ obeys the relation $Rv^{1/2}=$ const, rather than the DLA case of $Rv^0=$ const. This led to a relationship between models with different cutoffs and different $\eta$ such that the dominant growth probability was the same. The model with fixed $R$ that grows in the same manner as in fluid flow corresponds to $\eta \approx 1.2$. We were very encouraged in the channel by this analogy because the average of DBM with this $\eta$ did fit a Saffman-Taylor finger with $\lambda=1/2$ rather well. 

In the wedge, as Figures~(\ref{90}) and (\ref{60}) show, the DBM fingers \emph{do not fit the analytic shape}. They tend to be too narrow. This is exactly the opposite of the situation in the channel. Once more, there are no adjustable parameters available to us to fit. 

\section{Summary and discussion}
When we began this study we were confident that we would be able to put the averaging conjecture on a firm footing, based on our experience with the channel geometry. Our expectations were not at all fulfilled. It is possible to maintain that we have, in fact, eliminated the averaging conjecture altogether. Perhaps this strong conclusion is premature, but, certainly, the situation is not very clear. To summarize: in a channel, averaged DBM using  the mapping of Ref. \cite{ball02prl,ball03pre} gives a Saffman-Taylor finger with the  correct $\lambda$ but averaged DLA does not. In the wedge, averaged DLA gives the correct finger opening angle for both wedges that we looked at, but averaged DBM does not.

We might be tempted to say that our proposal for averaging based on conformal maps should simply be discarded. If we do that, we are reduced to using density averages which don't fit the analytic fingers any better (worse, in fact) and are ambiguous to boot.

In passing, we should comment on the situation with tip-splitting. We do not agree that DLA averages in a wedge should tip-split for any opening angle $> 0$. We base this on our  work in the wedge geometry \cite{kessler98}. In that paper we used not reflecting boundary conditions, but periodic boundaries for the wedge. We looked at angular correlations of the density for DLA clusters, and found that for small wedge angles there was a minimum in the correlation function half-way between the branch and its image. We interpreted this by saying that for small angles there was one major branch. For large angles, $\gtrsim$ 144$^\circ$ we found a secondary maximum in the correlation function, i.e. more than one major branch. These results do not directly carry over to the present case, but, qualitatively, we think that tip-splitting in the sense just described is not at all clearly present in this case. There is a numerical result of Ref. \cite{arneodo96} for a 60$^\circ$ wedge which seems to contradict this, but we are confident that our statistics were much better. 

It is likely that tip-splitting for DLA is a probabilistic matter. We suspect that in a 90$^\circ$ wedge some clusters split, but the majority do not. This could account for the small, but definite disagreement between the shape of the tip of the averaged clusters in the 90$^\circ$ wedge with the analytic solution. However, there is almost certainly a qualitative difference between DLA and fluid flow with respect to tip-splitting. Also, the extent  of tip-splitting found in mean-field theory \cite{levine92,arneodo96} does not agree with our DLA averages, or with the results of \cite{kessler98}. 

The project of formulating a description of the average over the DLA ensemble still seems to us to be quite a worthy one, However, the present results show that the current state of the art in this area is far from giving the definitive answer.

\section*{Acknowledgments}  We are indebted to Dave Kessler for the suggestion of measuring the average conformal map. We thank the Center for the Study of Complex Systems at the University of Michigan for computing resources. LMS acknowledges partial support by NSF grant No. DMS-0244419. ES would like to thank the University of Michigan for hospitality, 
and the PHYNECS training network of the European Commission for financial 
support (contract HPRN-CT-2002-00312).

 \bibliography{chaosdlarefs}
 \end{document}